\documentclass[conference]{IEEEtran}
\IEEEoverridecommandlockouts
\usepackage{cite}
\usepackage{amsmath,amssymb,amsfonts}
\usepackage{algorithmic}
\usepackage{graphicx}
\usepackage{textcomp}
\usepackage{xcolor}
\usepackage{soul}
\usepackage{hyperref}

\usepackage{braket}
\usepackage{amsmath,amsbsy,amsfonts,amssymb,amsthm,commath}

\def\ddefloop#1{\ifx\ddefloop#1\else\ddef{#1}\expandafter\ddefloop\fi}
\def\ddef#1{\expandafter\def\csname bb#1\endcsname{\ensuremath{\mathbb{#1}}}}
\ddefloop ABCDEFGHIJKLMNOPQRSTUVWXYZ\ddefloop
\def\ddef#1{\expandafter\def\csname c#1\endcsname{\ensuremath{\mathcal{#1}}}}
\ddefloop ABCDEFGHIJKLMNOPQRSTUVWXYZ\ddefloop
\def\ddef#1{\expandafter\def\csname v#1\endcsname{\ensuremath{\boldsymbol{#1}}}}
\ddefloop ABCDEFGHIJKLMNOPQRSTUVWXYZabcdefghijklmnopqrstuvwxyz\ddefloop
\def\ddef#1{\expandafter\def\csname v#1\endcsname{\ensuremath{\boldsymbol{\csname #1\endcsname}}}}
\ddefloop {alpha}{beta}{gamma}{delta}{epsilon}{varepsilon}{zeta}{eta}{theta}{vartheta}{iota}{kappa}{lambda}{mu}{nu}{xi}{pi}{varpi}{rho}{varrho}{sigma}{varsigma}{tau}{upsilon}{phi}{varphi}{chi}{psi}{omega}{Gamma}{Delta}{Theta}{Lambda}{Xi}{Pi}{Sigma}{varSigma}{Upsilon}{Phi}{Psi}{Omega}{ell}\ddefloop

\def\BibTeX{{\rm B\kern-.05em{\sc i\kern-.025em b}\kern-.08em
    T\kern-.1667em\lower.7ex\hbox{E}\kern-.125emX}}
\begin{document}

\title{Quantum Generative Models for Small Molecule Drug Discovery}
\IEEEaftertitletext{\vspace{-1\baselineskip}}

\author{\IEEEauthorblockN{Junde Li}
\IEEEauthorblockA{
\textit{Pennsylvania State University}\\
jul1512@psu.edu}
\and
\IEEEauthorblockN{Rasit Topaloglu}
\IEEEauthorblockA{\textit{IBM} \\
rasit@us.ibm.com}
\and
\IEEEauthorblockN{Swaroop Ghosh}
\IEEEauthorblockA{
\textit{Pennsylvania State University}\\
szg212@psu.edu}

}
\IEEEaftertitletext{\vspace{-1.5\baselineskip}}

\maketitle

\vspace{-10mm}
\begin{abstract}

Existing drug discovery pipelines take 5-10 years and cost billions of dollars. Computational approaches aim to sample from regions of the whole molecular and solid-state compounds called chemical space which could be on the order of $10^{60}$.
Deep generative models can model the underlying probability distribution of both the physical structures and property of drugs and relate them nonlinearly. By exploiting patterns in massive datasets, these models can distill salient features that characterize the molecules. 
Generative Adversarial Networks (GANs) discover drug candidates by generating molecular structures that obey chemical and physical properties and show affinity towards binding with the receptor for a target disease. However, classical GANs 
cannot explore certain regions of the chemical space and
suffer from curse-of-dimensionality. 
A full quantum GAN may require more than 90 qubits even to generate QM9-like small molecules. We propose a qubit-efficient quantum GAN with a hybrid generator (QGAN-HG) to learn richer representation of molecules via searching exponentially large chemical space with few qubits more efficiently than classical GAN. 
The QGAN-HG model is composed of a hybrid quantum generator that supports various number of qubits and quantum circuit layers, and, a classical discriminator. QGAN-HG with only 14.93\% retained parameters can learn molecular distribution as efficiently as classical counterpart.
The QGAN-HG variation with patched circuits considerably accelerates our standard QGAN-HG training process and avoids potential gradient vanishing issue of deep neural networks. Code is available on GitHub \href{https://github.com/jundeli/quantum-gan}{https://github.com/jundeli/quantum-gan}.

\end{abstract}


\section{Introduction}
\label{sec:intro}

Drug development pipeline consists of stages of target discovery, molecular design, preclinical studies, and clinical trials, which makes the process of creating a marketable drug expensive and time consuming \cite{gnc}. The majority of new drugs approved by US Food and Drug Administration is small-molecule drugs whose structural and functional diversity make their matching with biological binding sites possible \cite{beall2019pre}. Searching new drugs can be considered as navigating in the chemical space, which is the ensemble of all organic molecules, and navigation in unknown chemical space falls within the field of \textit{de novo} drug design \cite{reymond2012}.

\begin{figure*}
\centering
\includegraphics[width=17.5cm]{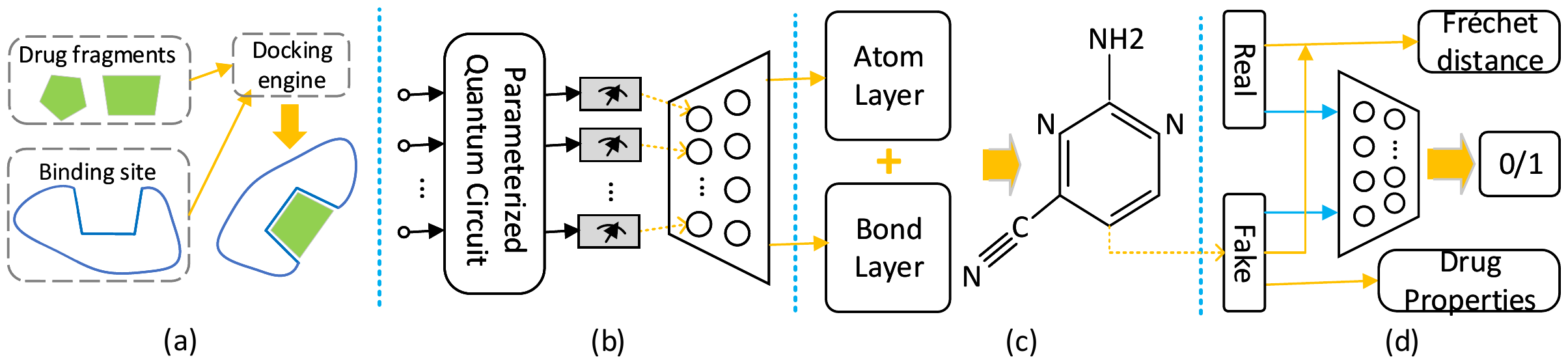}
\vspace{-3mm}
\caption{(a) Only generated molecules that have high affinity towards the receptor binding sites are considered as valid; (b) quantum stage (which is a parameterized quantum circuit with last-layer N measuring the expectation values) and classical stage (neural network with last-layer out-feature dimension of 512 \cite{molgan}) separated by blue dotted line; (c) application of atom layer and bond layer for generating synthetic molecular graphs (one example synthetic molecule is given); (d) a batch of real molecules from training dataset (QM9 in this case) and a batch of synthetic molecules generated from (c) are fed into classical discriminator for real/synthetic prediction and FD score calculation, and drug properties for synthetic molecules are evaluated using RDKit package \cite{rdkit}. The prediction losses from discriminator are back-forwarded to two neural networks as well as quantum circuit for updating all parameters simultaneously in each training epoch.}
\label{flow}
\vspace{-4mm}
\end{figure*}

Machine learning techniques have been explored in all development stages, especially molecular design with desirable properties \cite{ekins2019, batool2019, gnc}. Generative models such as, variational autoencoders (VAEs) \cite{kingma2013auto}, generative adversarial networks (GANs) \cite{gan} and recurrent neural networks (RNNs) are specifically adopted for learning latent representations of molecules and generating large amount of drug candidates for further high-throughput screening. Deep generative models have been used for various representation types of molecules such as, string-based, graph-based and shape/structure-based \cite{organ, kusner17jm, molgan, skalic2019target, skalic2019} representations. Generative learning with graph-structured molecules is invariant to the orderings of atoms \cite{molgan, simonovsky2018} and automates the navigation to a chemical region abundant in desired molecules. However, classical generative models cannot generate all possible distributions indicating that it cannot explore certain regions of the chemical space. 
This is primarily due to exponential choices to gradually add a new molecule in the existing drug fragment. Note, drug discovery can be explained using lock and key model where the receptor (a protein binding site associated with a disease) is considered as a lock and the drug is a key (Fig. \ref{flow}(a)). If the shape and pose of the drug is right, it can plug into the binding site curing the disease. 

Quantum computing can offer unique advantages over classical computing in many areas such as, chemistry simulation, machine learning, and optimization \cite{biamonte2017quantum, mcardle2020, nannicini2019}. Quantum GAN is one of the main applications of near-term quantum computers due to its strong expressive power in learning data distributions even with much less parameters compared to classical GANs \cite{stein2020qugan}. Quantum GANs can offer several opportunities e.g., (i) quantum speedup in the runtime making it possible to learn richer representation of molecules via deeper models due to the amplitude amplification property; (ii) ability to search exponentially large chemical space with few qubits and sample from distributions that may be difficult to model classically. 

Quantum GAN is still at its nascent stage due to qubit constraints on noisy quantum computers. 
Huang et al. \cite{pan} proposed a quantum patch GAN mechanism to efficiently use limited qubits for generating hand-written digit images, however, the method only suits two digits of 0 and 1. QuGAN \cite{stein2020qugan} aims to learn MNIST data set, but the original 784-dimension images were reduced to only 2 dimensions. \textit{To the best of our knowledge, no existing quantum GAN mechanisms can essentially solve real-world complicated learning tasks}. 

Drug molecules can be represented as graphs where the nodes and edges correspond to atoms and bonds, respectively. Given the task complexity of learning molecule distribution, full quantum GAN can hardly encode all training data in a quantum way. Take the small molecule dataset QM9 \cite{qm9} for example. The total number of qubits required for reconstructing synthetic molecules is $\binom{9}{2}\log5 + 9\log5 > 90$ where 5 is the number of bond types and atom types contained in QM9. No commercially available quantum computers support over 90 qubits at present. The proposed quantum GAN model can overcome the qubit constraint while still exploiting the benefits of quantum computing.

We propose a qubit-efficient quantum GAN mechanism with hybrid generator and classical discriminator for efficiently learning molecule distributions based on classical MolGAN \cite{molgan}. Since the proposed quantum GAN requires less qubits, simulation is still a viable option for training unlike full quantum GAN with large number qubits that cannot be simulated using classical computers. We also examine the patched circuit idea \cite{pan} by comparing to the original single large generator circuit implementation using metric of Fréchet distance and drug property scores. Fig. \ref{flow} shows the overall workflow of our qubit-efficient hybrid quantum GAN model for drug discovery. \textit{This work aims to discover a library of novel and valid molecules that can be screened by the docking engine in the next step (Fig. \ref{flow}(a)).}

\textbf{Novelty/contributions:} \textit{To the best of our knowledge, this is the first work on drug discovery using quantum machine learning}. We, (1) propose a novel quantum GAN mechanism with hybrid generator to qubit-efficiently tackle any real-world learning tasks solvable on classical GANs; 
(2) generate potentially better drug molecular graphs in terms of Fréchet distance (FD) score and drug property scores, and achieve high training efficiency by reducing generator architecture complexity; (3) examine the performance of patched quantum GAN for drug discovery task and comparison is made with QGAN-HG with a single quantum circuit; (4) validate the capability of generating small drug molecules on real IBM quantum computers by running inference stage of QGAN-HG model (not for training stage due to long waiting time on IBM Q machines per iteration).



\section{Background}
\label{sec:background}


\subsection{Computational Drug Discovery}

Generative models such as, GAN \cite{gan} have been explored for discovery of drug molecules with desired properties by learning drug molecule distribution based on given chemical dataset and sample synthetic molecules from the chemical space. GAN consists of two networks, namely generator (G) and discriminator (D), competing with each other. The generator uses noise as input to generate synthetic data sample whereas the discriminator flags if given sample is real or fake with a binary classifier. Generator $G(\vz; \theta_g)$ maps random input noise $z$ to synthetic chemical data space $p_g$, while discriminator $D(\vx; \theta_d)$ outputs a single scalar indicating the probability that $\vx$ come from real data rather than $p_g$. D is trained to maximize the probability to assign correct label and G is trained to minimize the difference between real and fake $\log(1-D(G(\vz)))$. The two-player minimax game is trained based on the following value function:
\begin{align*}
   \min_{\theta_g}\max_{\theta_d}V(D,G) &= \mathbb{E}_{\vx \sim p_{data}(\vx)}[\log D(\vx)] \\
   & + \mathbb{E}_{\vz \sim p_{z}(\vz)}[\log(1-D(G(\vz)))]
\end{align*}

Chemical compounds can be represented as graphs with nodes and edges designating various atoms and their bonds, respectively. For example, Fig. \ref{backg}(a-b) shows a molecule and its graph representation where atom types of N and O are encoded as 2 and 3 in atom vector, and bond types of single and double are encoded as 1 and 2 in bond matrix. If the generated structure is chemically stable and exhibits high affinity towards the receptor binding sites then it can be treated as a valid drug molecule.   The generator and discriminator are trained using example drugs/molecules until the synthetic chemical distribution is close to real chemical space. The quality of GAN outcome are measured by Fréchet distance and RDKit (for chemical properties) \cite{rdkit}.

\subsection{Quantum Machine Learning}

Quantum systems have atypical patterns that classical computers cannot produce efficiently \cite{biamonte2017quantum}. Machine learning tasks are sometimes hard to train on classical computers due to large-scale and high-dimensional data set. 
Quantum neural network (QNN) can represent given dataset, either quantum or classical, and be trained using a series of parameter dependent unitary transformations. QNN architecture is dependent on qubit count, quantum circuit layer, and quantum gates applied because the architecture is essentially a variational quantum circuit. Thus, below quantum computing concepts are helpful in understanding quantum neural network.

\begin{figure}
\centering
\includegraphics[width=8.5cm]{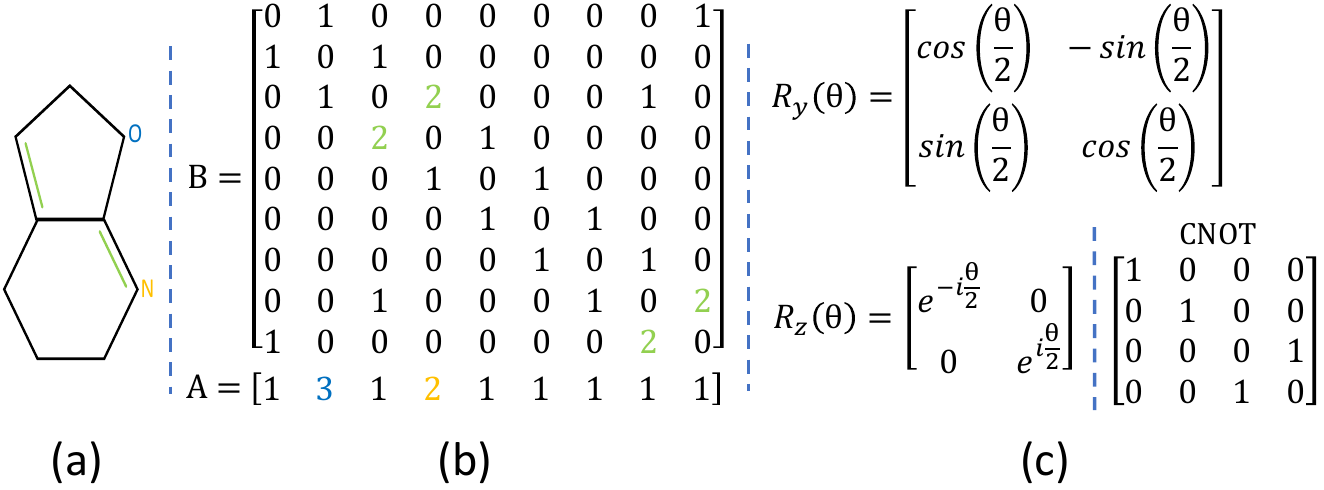}
\vspace{-3mm}
\caption{(a-b) A sample molecular graph from QM9 denoted by its corresponding atom vector A and bond matrix B; (c) all quantum gates used in this study.}
\label{backg}
\vspace{-4mm}
\end{figure}

\textbf{Quantum Circuit:} Quantum circuits consist of gates that modulate the state of the qubits to perform computation. Gate pulses induce a varied amount of rotation along the axes in the Bloch sphere. Quantum gates could be applied on 1 qubit (e.g., rotation gates $R_y$ and $R_z$) or 2 qubits (e.g., CNOT gate), as shown in Fig. \ref{backg}(c).
Finally, measurement is applied for getting the expectation value after certain number of shots.

\textbf{Quantum Noise:} Noisy quantum computers suffer from noise sources such as, T1 relaxation time, T2 dephasing time, gate errors and readout error. These are also called qubit quality metrics. Crosstalk, qubit-to-qubit variation and temporal variations in qubit quality also exist. 
Fortunately, reasonable quantum noise level is not detrimental for QNN, rather it can even be beneficial because systematic noise is helpful for improving generalization performance of neural networks. Instead of mitigating such noises, we rather explore the effect of quantum noise by running the inference for QGAN-HG on a real IBM quantum computer.

\section{Quantum Generative Adversarial Networks}
\label{sec:qgan_hg}


\subsection{Quantum GAN Flavors}

Quantum GAN has a few flavors of the generator and discriminator implementations depending on their execution environments, either on quantum computers, classical machines or quantum simulators. The flavor with quantum discriminator is not applicable here due to limited number of qubits on near-term quantum computers. Real data shown in Fig. \ref{flow}(d) has to engage the state preparation stage, usually through amplitude encoding, for encoding classical data in a quantum state, and this stage takes $N\log(M)$ qubits where $N$ is the training set size and $M$ is feature dimension \cite{pan, plesch2011}.
The flavor with a pure quantum generator is not directly applicable either considering the complicated task of drug discovery. As noted in Section~\ref{sec:intro}, more than 90 qubits are needed to discover QM9-like molecules (not suitable for near-term quantum computers). Thus we propose a new quantum GAN with hybrid generator and classical discriminator to exploit the quantum benefits.

\subsection{Quantum GAN with Hybrid Generator}

Quantum GAN with hybrid generator (QGAN-HG) is composed of a parameterized quantum circuit to get a feature vector of qubit size dimension, and a classical deep neural network to output an atom vector and a bond matrix for the graph representation of drug molecules. Another patched quantum GAN with hybrid generator (P-QGAN-HG) is considered as the variation of QGAN-HG where the quantum circuit is formed by concatenating few quantum sub-circuits.
 
\begin{figure}
\centering
\includegraphics[width=8cm]{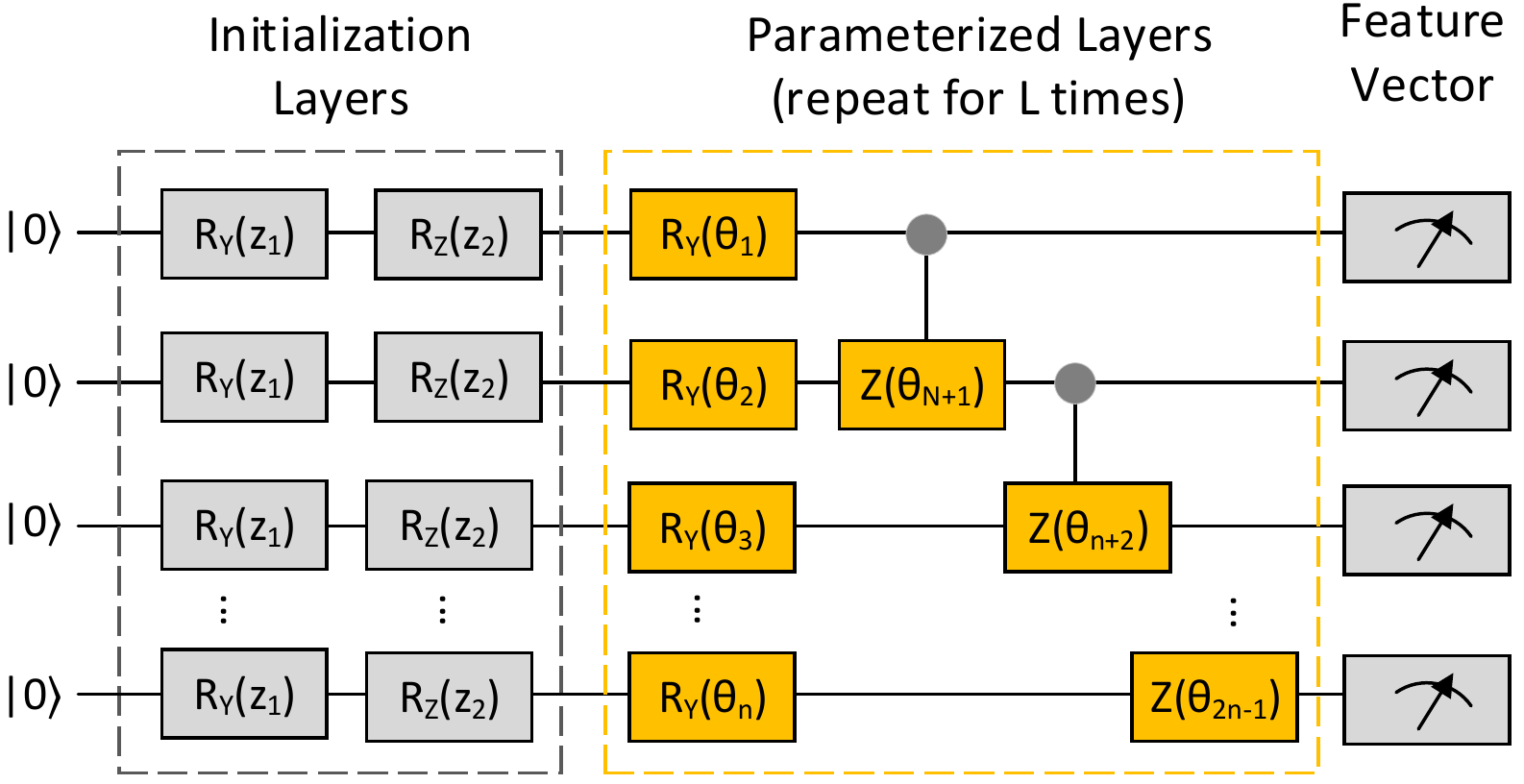}
\vspace{-3mm}
\caption{Parameterized quantum circuit to obtain feature vector of N dimensions. The circuit is composed of initialization layers, repeatable parameterized layers and measurement layer. Two CNOT gates for each ZZ interaction for creating entanglement are not shown here.}
\label{pqc}
\vspace{-4mm}
\end{figure}

\begin{figure*}
\centering
\includegraphics[width=17cm]{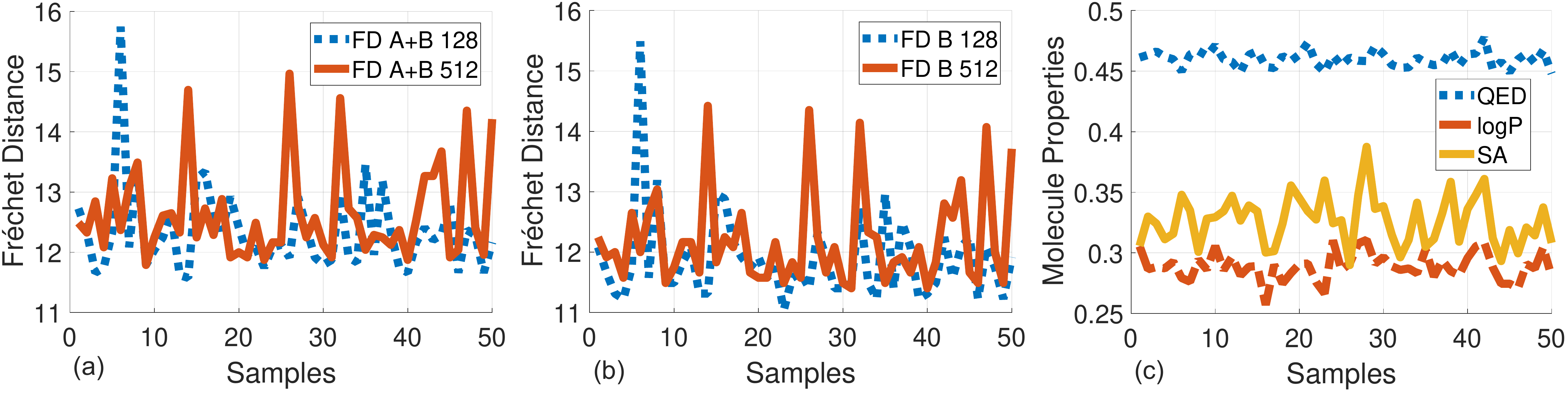}
\vspace{-3mm}
\caption{Metrics of Fréchet distance and molecule properties for real data points: (a) Fréchet distances calculated based on both Atom and Bond (A+B) with batches of 128 molecules and 512 molecules; (b) Fréchet distances calculated based on only Bond (B) for batch sizes of 128 and 512; (c) three major drug properties evaluated on sample batches of 128 molecules from (a). Note that these 50 samples are independently sampled from QM9.}
\label{metric}
\vspace{-4mm}
\end{figure*}

\textbf{QGAN-HG Quantum Circuit:} In this variant, a quantum layer is added for exploiting the strong expressive power of variational quantum circuits which perform low-rank matrix computations in $\mathcal{O}(poly(\log(M)))$ time for exponential speedup \cite{pan, lloyd2018quantum}. The variational quantum circuit (Fig. \ref{pqc}) consists of 3 stages, namely initialization, parameterized (repeatable for L layers with $L(2N-1)$ parameter count) and measurement stages. Two parameters $z_1$ and $z_2$ are uniformly sampled from $[-\pi, \pi]$, which essentially substitute the random Gaussian noise input for classical GANs. After applying the initialization layers, the input state in mathematical form $\ket{z} = (\otimes_{i=1}^NR_Z(z_2)R_Y(z_1)\ket{0})^{\otimes^N}$ is prepared. Let us denote the parameterized layers repeated for $L$ times as unitary matrix $U(\vtheta)$. The final quantum state is of the form $\ket{\Psi(z)} = U(\vtheta)\ket{z}$.
A series of measurement operators are applied to obtain the expectation value for each qubit and further form the feature vector to be fed to classical neural network.

\textbf{QGAN-HG Neural Network:} This classical stage of hybrid generator is a standard neural network with input layer receiving the feature vector of expectation values. The final layer consists of the separate atom and bond layers for creating atom vectors and bond matrices, respectively. Like MolGAN \cite{molgan}, a categorical re-parameterization step with Gumbel-Softmax \cite{jang2016}, which supports gradient calculation in the backward pass, is taken for getting discrete fake molecular graphs. Note that, 85.07\% and 98.03\% of generator parameters are respectively cropped by reducing major linear layers from classical GAN \cite{molgan} so as to demonstrate the strong expressive power of quantum circuits. Due to the necessity of reconstructing QM9-like molecules (with structure of $\vX \in \mathbb{R}^{9X5}$ for atoms and $\vA \in \mathbb{R}^{9X9X5}$ for bonds), neural network architecture can hardly be further reduced.

\textbf{Patched QGAN-HG:} The patched quantum GAN with hybrid generator consists of the same two ingredients as above. However, the parameterized quantum circuit consists of multiple quantum sub-circuits. Theoretically, P-QGAN-HG has its pros and cons relative to QGAN-HG with an integral quantum circuit. P-QGAN-HG requires less quantum resources because multiple sub-circuits can be executed sequentially or in parallel. Another benefit is that each circuit can be simulated more efficiently, which speeds up the learning process accordingly. However, one of the obvious drawbacks is reduced expressive power since quantum state dimension is reduced from $2^N$ to $2^{N/2}$ (say two circuits with half size) in Hilbert space. The performance of patched QGAN-HG is compared with QGAN-HG in the following section.

\textbf{Discriminator and Optimizer:} The discriminator is kept the same as MolGAN \cite{molgan} as its parameter size is at par with the hybrid generator. However, the reward network is discarded since reward value is too minuscule to noticeably contribute to training the model. Generated molecules are evaluated using RDKit together with Fréchet distance based metric. Quantum gate parameters and weights in neural network are updated simultaneously using a single optimizer, while discriminator uses a separate one for being updated alternatively.

\begin{figure*}
\centering
\includegraphics[width=17cm]{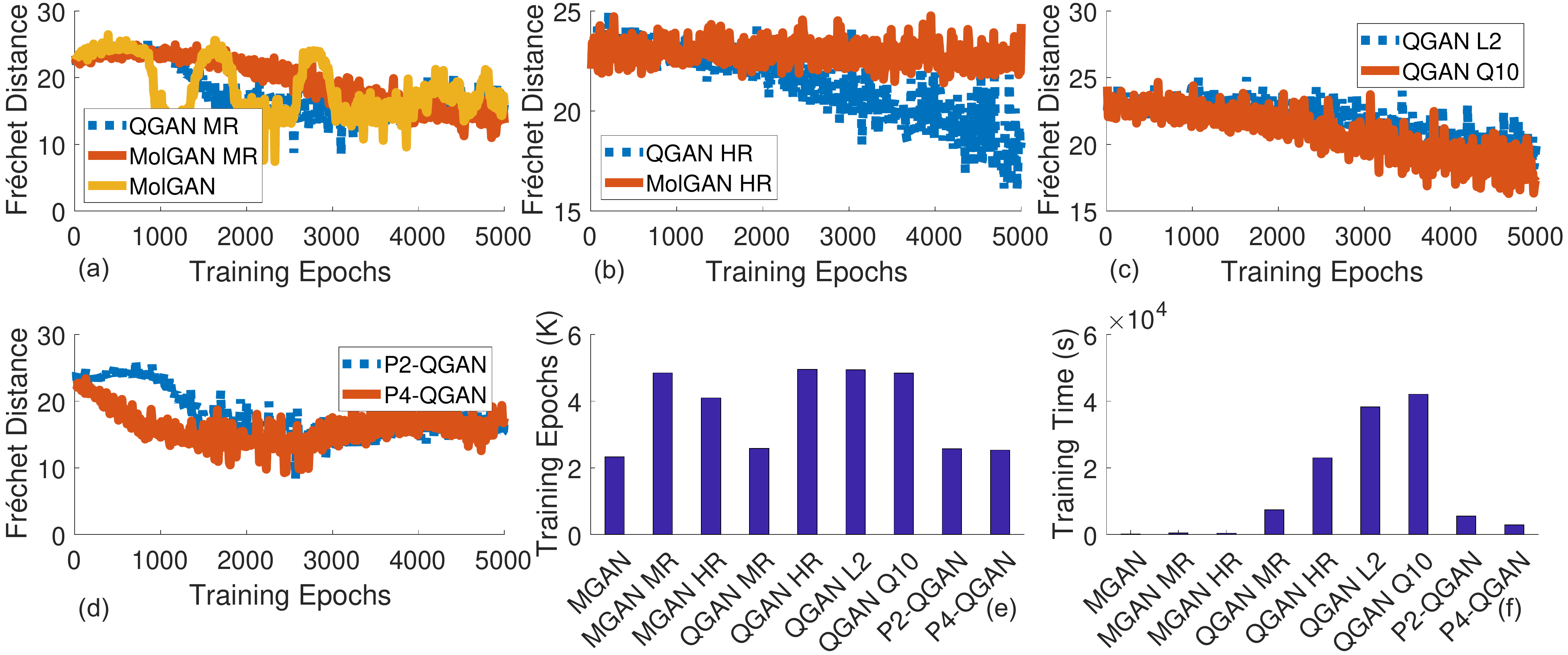}
\caption{Training comparison among GAN flavors: (a) Fréchet distances for MolGAN, moderately reduced (14.93\%) MolGAN and QGAN-HG; (b) Fréchet distances for highly reduced (1.97\%) MolGAN and QGAN-HG; (c) learning curves for highly reduced QGAN-HG with quantum circuit level $L=2$ and $N=10$, respectively; (d) learning curves for patched QGAN-HG with two sub-circuits (4 qubits for each) and four sub-circuits (2 qubits for each); (e) training epochs with lowest Fréchet distances for all GAN flavors; (f) training times elapsed (early stopping at epochs from (e)) for all GAN flavors.}
\vspace{-3mm}
\label{training}
\end{figure*}

\section{Experimental Setup}
\label{sec:experiments}


\subsection{Dataset and Metrics}

\textbf{Dataset:} All the experiments are conducted with quantum machine learning benchmarking QM9 \cite{qm9} dataset which contains 133,885 molecules with up to 9 heavy atoms of types of carbon, nitrogen, oxygen, and fluorine.

\textbf{Fréchet Distance:} Learning results of the proposed GANs are evaluated with Fréchet distance metric which measures the similarity between real and generated molecule distributions. Generated molecule distribution is approximately created by generating a batch of molecules, and real one is approximately formed by randomly sampling the same number of molecules from QM9 dataset. Each sample batch of molecules is concatenated and considered as a multi-dimensional point in the distribution, then Fréchet distance is calculated using 50 of these points (sampled for 50 batches from both distributions). Fig. \ref{metric} shows Fréchet distances calculated from two batch sizes of 128 and 512 molecules and independently sampled for 50 times. FD A+B (see Fig. \ref{metric}(a)) denotes the similarity calculated based on both atom and bond matrices; while FD B (see Fig. \ref{metric}(b)) only on bond matrices. FD A+B includes more random noises from molecule atoms and is projected to severely disturb the similarity between real molecule batches. Interestingly, FD A+B correlates well with the distance calculated without atoms. This is a strong evidence of the inherent connection between atoms and bonds. The means (12.3342, 12.6387) and variations (0.7057, 0.7849) between FD calculated for 128 and 512 batch sizes are close. Thus, all following experiments are evaluated with FD A+B metric and 128 batch size.

\textbf{Drug Properties:} Molecule properties are the metrics for drug quality evaluation during inference stage. Three primary properties include, (i) quantitative estimate of druglikeness (QED) which measures the likelihood of compound being a drug; (ii) log octanol-water partition coefficient (logP) which measures the solubility of a compound; and (iii) and synthetic accessibility (SA) which quantifies the ease of a compound being synthesized in pharmaceutical factory. Together with other properties, they are measured using RDKit.

\subsection{Implementation Details}

The quantum circuits can be executed either on a simulator or real quantum machine. The simulator supports customized setting of noise levels and sources (noiseless environment is set in this paper), while real quantum devices have different noise characteristics across different machines.

\textbf{Training:} We pivot on the classical MolGAN \cite{molgan} to implement our QGAN-HG and P-QGAN-HG algorithms. As mentioned in Section~\ref{sec:qgan_hg}, some linear layers and reward network are removed in our experiments for evaluating the expressive power of quantum circuit and using drug property metrics fairly. Our QGAN variations are trained with a mini-batch of 128 molecules using Adam optimizer on a single RTX 2080 Ti GPU for classical part and PennyLane platform \cite{pennylane} with default qubit plugin for the quantum stage. As explained in Section~\ref{sec:intro}, real quantum machine is not utilized during training stage due to long waiting time on IBM Q machines. The learning rate is initially set to 0.0001 for both generator and discriminator and starts decaying uniformly at a factor of $1/2000$ after 3000 epochs. Total training epoch is set with 5000, and early stopping based on Fréchet distance is applied if model collapse happens.

\textbf{Inference:} Only hybrid generator is involved during inference stage. Since QGAN is well trained, the quantum circuit in the generator is executed on both PennyLane simulator and real quantum device of IBM Q16 Melbourne for comparison. Drug quality for generated molecules are evaluated by properties such as, QED, logP, and SA, among others, all of which are normalized to be within $[0, 1]$. Finally, GAN variations are compared by taking 1000 generated molecules.

\section{Evaluation Results}
\label{sec:evaluation}


\subsection{QGAN-HG Results}

\begin{table*}[!ht]
\fontsize{10pt}{10pt}
\centering
\caption{Drug properties of 1000 generated molecules from all GAN variations in this paper. Best results are shown in bold. 
Sign '-' indicates the corresponding metric for sampled molecules are not successfully evaluated by RDKit.} 

\begin{tabular}{|c| c| c| c| c| c| c| c| c| c| c|}
 \hline
 Method & Druglikeliness & Solubility & Synthesizability & Diversity & Valid & Unique & Novel \\
 
 \hline  \hline 
 MolGAN$^*$ \cite{molgan} &  0.50 & \textbf{0.70} & 0.11 & \textbf{1.0} & \textbf{0.82} & 0.21 & \textbf{1.0} \\
 MolGAN MR   &  0.47 & 0.60 & 0.14 & \textbf{1.0} & 0.31& 0.70 & \textbf{1.0} \\
 MolGAN HR  &  - & - & - & \textbf{1.0} & 0.10 & \textbf{1.0} & \textbf{1.0} \\
 QGAN-HG MR (proposed)  &  \textbf{0.51} & 0.49 & 0.07 & \textbf{1.0} & 0.63 & 0.35 & \textbf{1.0} \\
 QGAN-HG HR (proposed)  &  - & - & - & \textbf{1.0} & 0.03 & \textbf{1.0} & \textbf{1.0} \\
 QGAN-HG HR L2 (proposed)  &  - & - & - & \textbf{1.0} & 0.02 & \textbf{1.0} & \textbf{1.0} \\
 QGAN-HG HR Q10 (proposed)  &  0.49 & 0.43 & 0.15 & \textbf{1.0} & 0.04 & \textbf{1.0} & \textbf{1.0} \\
 P2-QGAN-HG MR (proposed) &  0.49 & 0.62 & 0.11 & \textbf{1.0} & 0.53 & 0.40 & \textbf{1.0} \\
 P4-QGAN-HG MR (proposed) &  0.49 & 0.51 & 0.13 & \textbf{1.0} & 0.59 & 0.45 & \textbf{1.0} \\
 QGAN-HG MR (on IBM quantum computer) &  0.48 & 0.50 & \textbf{0.17} & \textbf{1.0} & 0.38 & 0.92 & \textbf{1.0} \\
 \hline
 
  \multicolumn{8}{l}{MolGAN$^*$ refers to MolGAN \cite{molgan} trained in the present study.}
  
 \end{tabular}
\label{properties}
\vspace{-4mm}
\end{table*}

Fig. \ref{pqc} shows QGAN-HG performance may rely on both qubit count $N$ and repeatable layer count $L$. Higher qubit count and layer count presumably correspond to stronger expressive power of hybrid generator. To demonstrate expressive power, we reduce the neural network parameter count to two levels, namely, 14.93\% (\textbf{MR}-moderately reduced) and 1.97\% (\textbf{HR}-highly reduced) of generator parameters of original MolGAN. Fig. \ref{training}(a-b) show the training performance comparison between MolGAN and QGAN-HG for moderately and highly reduced architectures, respectively. One can observe from Fig. \ref{training}(a) that all mechanisms can reach a reasonably good training point (see Fig. \ref{metric} for benchmark) within 5000 epochs, however, moderately reduced MolGAN takes around 4000 iterations while original MolGAN and QGAN-HG take only 2500 iterations or so. Also note that, MolGAN and QGAN-HG both reach a slightly lower Fréchet distance than the reduced classical counterpart. As shown in Fig. \ref{training}(b), MolGAN with highly reduced architecture can hardly be learned though a slight downward trend is observed. The weak learning ability of MolGAN-HR is attributed to mainly two reasons: (1) the features of QM9 drug molecules cannot be well represented using a light-weight neural network; (2) parameter count of generator is not at par with that of discriminator. Intriguingly, a sharp downward learning curve for QGAN-HG is still observed. It is worth mentioning that only 15 gate parameters are involved in the quantum circuit. These are clear evidences of strong expressive power of variational quantum circuits.

Model collapsing occur for GAN variations if Fréchet distances (approximate indicator of learning quality) start increasing after certain training point. We adopt early stopping technique (the lowest reached point of Fréchet distance) to somewhat prevent the training instability issue in GANs. To measure the effects of circuit layer and qubit count, we also implement QGAN-HG with $L=2$ and $N=10$ separately. However, the enhanced QGAN-HG variations with more circuit layer and qubits (see Fig. \ref{training}(c, e)) do not help accelerate learning process much relative to QGAN-HG in Fig. \ref{training}(b).

\subsection{P-QGAN-HG Results}

The patched QGAN-HG mechanism is developed on the basis of \cite{pan}, however, P-QGAN-HG we proposed here uses all qubits for creating feature vector and has no specific qubits for non-linear mapping because of the following classical neural network. We demonstrate the expressive power of two patched QGAN-HG variations, i.e. P2-QGAN with two sub-circuits (each has 4 qubits and 7 gate parameters) and P4-QGAN with four sub-circuits (each has 2 qubits and 3 gate parameters). Surprisingly, the learning quality of these patched QGANs are comparable to QGAN with an integral circuit, as shown in Fig. \ref{training}(a, d), though patched QGANs have even less gate parameters. Further, the simulation time (see Fig. \ref{training}(f)) for patched quantum circuits are significantly reduced because of smaller qubit count and early convergence. Therefore, we consider patched QGAN-HG with multiple sub-circuits is an alternative to classical GAN since GAN training issues such as instability and vanishing gradients can be mitigated by shortening neural network depth.

\subsection{Drug Properties}

The training of GAN variations is evaluated by Fréchet distance, whereas the quality of drug molecules generated from GANs is specifically evaluated using a series of molecule property metrics, three of which are visualized in Fig. \ref{metric}(c). The drug property evaluation is performed by a specific inference stage. All GAN variations pick a point with lowest Fréchet distance within 5000 epochs for inference. 
We also run the inference stage for QGAN-HG on IBM Q quantum machines as well. Drug properties are calculated using 1000 sampled molecules. Table \ref{properties} displays the drug property results which are generally consistent with Fréchet distance results. Note that diversity and novel scores for all models are high, whereas synthetic accessibility is low relative to benchmark shown in Fig. \ref{metric}(c). QGAN-HG results executed on simulator and real quantum computer are consistent except the latter shows higher unique and systhesizability scores.

\section{Conclusion}
\label{conclusion}

We propose a novel quantum GAN with a hybrid generator for discovery of new drug molecules. Our hybrid GAN with patched quantum circuits concatenates feature vectors from different patches. Results show that classical GAN with 85.07\% reduced parameters cannot properly learn molecule distribution, however our QGAN-HG with only 15 extra quantum gate parameters can learn molecular task at par with original GAN. The proposed QGAN-HG with 98.03\% reduced parameters is significantly efficient than the highly reduced classical counterpart. The patched quantum GAN achieves comparable learning accuracy in terms of drug properties with even only 12 extra quantum gate parameters, and considerably accelerates the quantum GAN with an integral quantum circuit. The proposed patched quantum GAN model can be alternative to classical GAN due to less training time and avoidance of possible gradient vanishing problem in deep neural networks.

\bibliographystyle{IEEEtran}
\bibliography{IEEEabrv,ref}

\end{document}